\documentclass[12pt]{article}
\usepackage{latexsym, epsfig, graphics}
\newcommand{\be}{\begin{equation}}
\newcommand{\ee}{\end{equation}}
\newcommand{\bq}{\begin{eqnarray}}
\newcommand{\eq}{\end{eqnarray}}
\begin{document}
\begin{titlepage}
\today          \hfill 
\begin{center}

\vskip .5in

{\large \bf Mean Field Method Applied To The New World Sheet Field
Theory: String Formation}
\footnote{This work was supported in part
 by the Director, Office of Science,
 Office of High Energy  Physics, 
 of the U.S. Department of Energy under Contract 
DE-AC02-05CH11231.}

\vskip .50in


\vskip .5in
Korkut Bardakci
\footnote{Email:kbardakci@lbl.gov}
\vskip 9pt
{\em Department of Physics\\
University of California at Berkeley\\
   and\\
 Theoretical Physics Group\\
    Lawrence Berkeley National Laboratory\\
      University of California\\
    Berkeley, California 94720}
\end{center}

\vskip .5in

\begin{abstract}

The present article is based on a previous one, where a second quantized
field theory on the world sheet for summing the planar graphs of
$\phi^{3}$ theory was developed. In this earlier work, the ground state of the
model was determined using a variational approximation. Here, starting with
the same world sheet field theory, we instead use the mean field method
to compute the ground state, and find results that are in agreement with
the variational calculation. Apart from serving as a check on the variational
calculation, the mean field method enables us to go beyond the ground state
to compute the excited states of the model. The spectrum of these states is that
of a string with linear trajectories, plus a continuum that starts at
higher energy. We show that, by
 appropriately tuning the parameters of the model, the string spectrum
can be cleanly seperated from the continuum.

\end{abstract}
\end{titlepage}

\newpage
\renewcommand{\thepage}{\arabic{page}}
\setcounter{page}{1}
\noindent{\bf 1. Introduction}
\vskip 9pt

This paper is a natural follow up to a previous work [1], where a new
approach to the world sheet descripton of the planar graphs of the
$\phi^{3}$ field theory
was formulated. In contrast to the earlier work on the same subject
[2,3,4,5], which used a  first quantized formalism, the new formulation 
is based on second quantization on the world sheet.
 We have argued in [1] that this new
formulation is both simpler and better founded than the old one. In the
same reference,  using a variational ansatz, an approximate ground state
of the second quantized Hamiltonian was constructed  in $5+1$ space-time
dimensions. The ground state energy
and the coupling constant turned out to be ultraviolet divergent, needing
renormalization. Reassuringly, these divergences were the ones expected
from the perturbation expansion of the underlying field theory.

In the present article, instead of the variational method,
we use the mean field approximation for the same second quantized
Hamiltonian. The motivation for doing this is twofold: We would like
to check on the variational results, using a different approximation
scheme. The mean field method has a long and honorable history in
various branches of physics, and it was used in some of the early work 
[3,4,5] on this problem. It is reassuring that, as we shall see,
using this alternate approach, we are able to confirm the results
obtained in [1].

The second reason for trying a different approximation scheme is connected
with the limitations of the variational method, which is useful only for
investigating the ground state of a quantum system.
For example, it is very difficult to extend its reach to the excited
states. In contrast, using the mean field method, one can study the
model in full generality, including the excited states. One of the
main goals of this work is to show that with suitable tuning of the
parameters of the model, there is string formation on the world sheet.
We establish this result by showing that the spectrum of the excited
states is that of a string.

As in the earlier work,
 a central role is played by the field $\rho$ defined on the world
sheet by eq.(7). Roughly, $\rho$ measures the density of Feynman
graphs on the world sheet, a concept which we will make more precise
later on. An important question is whether  $\rho_{0}$, the ground
state expectation value of $\rho$, is different from zero. $\rho_{0}$
vanishes in any finite order of perturbation theory, whereas a non-
zero value for  $\rho_{0}$ means that the world sheet is densely covered
by graphs, and the contribution of high (infinite) order graphs dominate.
This can be thought of as a new phase of the underlying field theory,
different from the perturbative phase. It is natural to expect that
a world sheet densely covered by graphs would lead to a Nambu type action
and hence result in string formation; in fact, this was the picture that
motivated some of the very early work on this subject [6,7]. One of the main  
results of this paper is to show that the mean field calculation gives
 a non-zero $\rho_{0}$. We will show that such a non-trivial
background exits as a solution to field equations, and furthermore
it minimizes the ground state energy.
 This background will turn out to be an essential first step
 to showing string formation.

Along the way, we will study the ultraviolet divergences that are present
in the expansion around the non-trivial background. These turn out the to be
the ones expected from perturbation theory, and they can be renormalized.
Of course, the divergences depend on the number of dimensions; in
addition to $5+1$ dimensions, where the field theory is renormalizable
and asymptotically free, we also consider $3+1$ dimensions, where the theory
is superrenormalizable. The reason for this
  is that the superrenormalizable
model with a fixed coupling constant is much easier to analyze, and provides
a good warm up exercise for the more interesting but more difficult case
of $5+1$ dimensions. We find it very encouraging that the mean field
approximation simultaneously captures two desirable complimentary
features: The ultraviolet behaviour is consistent with perturbation theory,
and in the infrared region, the highly non-perturbative phenomenon of
string formation takes place.

Let us summarize the main results to emerge from this investigation.
Applying the mean field approximation to the world world sheet field theory 
developed in [1], we have found a non-trivial solution to field equations,
with $\rho_{0}\neq 0$, which minimizes the ground state energy.
 This background is in agreement with the variational
wave function derived in [1]. It also means that the phase where the world
sheet is densely covered by graphs is energetically prefered. Expanding
 around this background to second order, and
with some tuning the parameters of the
model, we find that the spectrum of the model is that of a 
bosonic string.

In order to have a self contained paper, in sections 2 and 3, we review
briefly the background needed to understand the present work. In section 2,
we discuss the general setup on the world sheet [2,8], and in section 3, we
review the world sheet field theory developed in [1]. In section 4, we
develop the mean field approximation and set up the corresponding
mean field equations. In section 5, these equations are used to find 
the ground state of the model in both $3+1$ and $5+1$ dimensions. Along
the way, we encounter ultraviolet divergences and show how to eliminate
them by renormalization. In particular, in $5+1$ dimensions, the cutoff
dependence of the bare coupling constant is consistent with asymptotic 
freedom. In section 6, we expand the Hamiltonian to second order around
the background found in the previous section. The solution to the resulting
equations yields two different types of spectra: bound states at lower
energies and a continuum starting at a higher energy. In section 7,
we show that the spectrum of the
bound states is that of a string with linear trajectories; however, the
 higher excited states of the string mix with the continuum, which
 muddies the string picture. It turns out that, by a suitable
adjustment of the parameters of the model, one can push the continuum
arbitrarily high, and thereby extend the string picture to 
arbitrarily high energies. Finally, in section 8, we summarize our
conclusions and suggest some directions for future research.

\vskip 9pt

\noindent{\bf 2. The World Sheet Picture}

\vskip 9pt

The generic planar graphs of the $\phi^{3}$ in the mixed light cone
 representation of 't Hooft [8] have a particularly simple form. The
world sheet is parametrized by the two coordinates
$$
\tau = x^{+} =(x^{0} + x^{1})/\sqrt{2},\,\,\sigma= p^{+} =
(p^{0} + p^{1})/\sqrt{2}.
$$
A general planar graph is represented by a collection of horizontal solid
lines (Fig.1), where the n'th solid line carries a $D$ dimensional
transverse momentum ${\bf q}_{n}$.
\begin{figure}[t]
\centerline{\epsfig{file=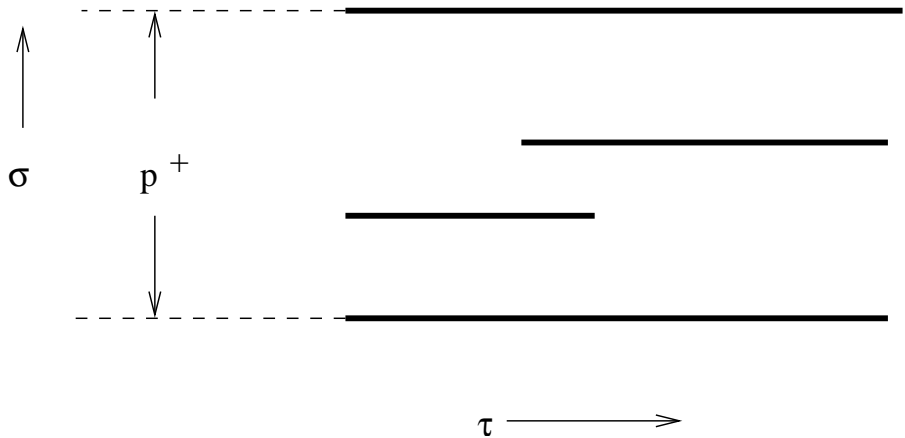, width=12cm}}
\caption{A Typical Graph}
\end{figure}
 Two adjacent solid lines labeled by
n and n+1 correspond to the light cone propagator
\be
\Delta(p_{n})=\frac{\theta(\tau)}{2 p^{+}}\,\exp\left(-i\tau\,\frac
{{\bf p}_{n}^{2}+m^{2}}{2 p^{+}}\right),
\ee
where ${\bf p}_{n}={\bf q}_{n}-{\bf q}_{n+1}$. A factor of $g$, the 
coupling constant, is inserted at the beginning and at the end of 
each line, where the interaction takes place.

For technical reasons, it is convenient to discretize the coordinate
$\sigma$ in steps of length $a$, which amounts to compactifying the
light cone coordinate $x^{-}=(x^{0} - x^{1})/\sqrt{2}$ at radius
$R= 1/a$. This type of compactification was  introduced by
Casher [9] in the context of the lightcone quantization of gauge theories.
In the context of the lightcone worldsheet and the discretization of the
$\sigma$ coordinate to enable summing planar diagrams, this discretization
was first proposed and exploited by Giles and Thorn [10]. Later, it was
found useful in
connection with the M theory [11,12]. In this 
paper, the $\sigma$ coordinate will always be discretized; in contrast,
 the time coordinate $\tau$ will remain continuous.

We also have to specify the boundary conditions to be imposed on the
world sheet. For simplicity, the coordinate $\sigma$ is compactified
by imposing periodic boundary conditions at $\sigma=0$ and $\sigma=p^{+}$,
where $p^{+}$ is the total $+$ component of the momentum flowing
through the whole graph. In contrast, since we will adopt the Hamiltonian
approach, the boundary conditions at $\tau=\pm \infty$ will be left free.

There is a useful way of visualizing the discretized world sheet. As 
pictured in Fig.2, the world sheet consists of horizontal dotted and
solid lines, spaced a distance $a$ apart.
\begin{figure}[t]
\centerline{\epsfig{file=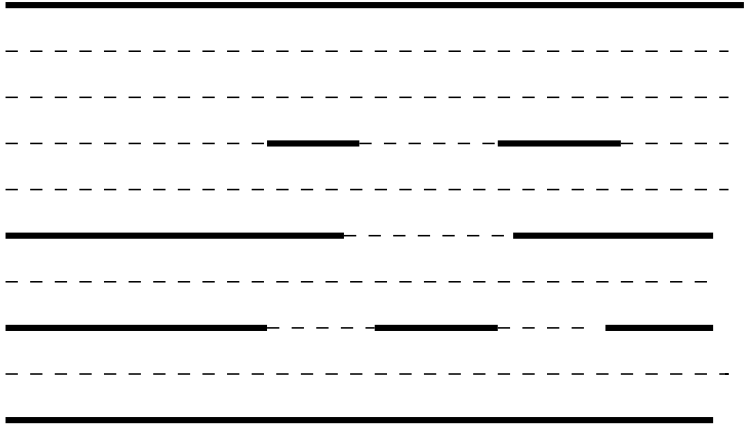, width=10cm}}
\caption{Solid And Dotted Lines}
\end{figure}
 Just as in Fig.1, the boundaries
of the propagators are marked by solid lines, in contrast,
  the bulk is filled with
dotted lines. Ultimately, one has to integrate over all possible
locations and lengths of solid lines, as well as over the transverse
momenta they carry. 

\vskip 9pt

\noindent{\bf 3. The World Sheet Field Theory}

\vskip 9pt

In this section, we will briefly review the world sheet field theory
developed in [1], which reproduces the light cone graphs described in the
previous section. We start by introducing the bosonic field
$\phi(\sigma,\tau,{\bf q})$, and its conjugate $\phi^{\dagger}(\sigma,
\tau,{\bf q})$, which at time $\tau$ respectively annihilate and create
a solid line carrying momentum ${\bf q}$ and located at site labeled by
$\sigma$. They satisfy the commutation relations
\be
[\phi(\sigma,\tau,{\bf q}),\phi^{\dagger}(\sigma',\tau',{\bf q'})]=
\delta_{\sigma,\sigma'}\,\delta({\bf q} -{\bf q'}).
\ee
The vacuum corresponds to a state with only dotted lines (empty world 
sheet), and it satisfies
\be
\phi(\sigma,{\bf q})|0\rangle =0.
\ee
 Since we are in the Hamiltonian picture with time $\tau$
is fixed, in equations of this type,
 we do not  usually explicitly write the time dependence. By
applying $\phi^{\dagger}$ s on vacuum, one can then construct states with
arbitrary number of solid lines.

Having defined the Fock space, a first go at the Hamiltonian could look like
the following:
\bq
H&=&H_{0} + H_{I},\nonumber\\
H_{0}&=& \sum_{\sigma'>\sigma}\int d{\bf q} \int d {\bf q'} \frac{
({\bf q}-{\bf q'})^{2} +m^{2}}{2 (\sigma' -\sigma)} \phi^{\dagger}(\sigma,
{\bf q})\, \phi(\sigma,{\bf q}) \, \phi^{\dagger}(\sigma',
{\bf q'})\, \phi(\sigma',{\bf q'}),\nonumber\\
H_{I}&=& g \sum_{\sigma} \int d{\bf q} \left(\phi(\sigma,{\bf q})+
\phi^{\dagger}(\sigma,{\bf q})\right).
\eq 
It is easy to check that
 $H_{0}$, applied to a state with two solid lines, generates
the free propagator of eq.(1), and $H_{I}$, converting a solid line into
a dotted one and vice versa, generates the interaction. There are, however,
several problems with this guess for the Hamiltonian:\\
a) The Hilbert space has redundant states,
corresponding to multiple solid lines at the
same site, generated by repeated
applications of $\phi^{\dagger}$ at the same $\sigma$.\\
b) Propagators should be assigned only to adjacent solid lines, whereas
the above $H_{0}$ generates unallowed propagators associated with
non-adjacent solid lines.\\
c) The prefactor $1/(2 p^{+})$ of the propagator is missing.

These problems can be solved simultaneously by introducing a two component
fermion field $\psi_{i}(\sigma,\tau)$, $i=1,2$, and its adjoint
$\bar{\psi}_{i}$ on the world sheet [13]. They satisfy the standard
anticommutation relations
\be
[\psi_{i}(\sigma,\tau),\bar{\psi}_{i'}(\sigma',\tau)]_{+}=\delta_{i,i'}
\delta_{\sigma,\sigma'},
\ee
and propagate freely on an uninterrupted line. The fermion with $i=1$
lives on the dotted lines and the one with $i=2$ lives on the solid
lines. It was shown in [1] how
to overcome the problems listed above with the help of the fermions.
Here, we will only present the final result, and refer the reader to
[1] for the details. 

To get rid of the redundant states, we impose the following constraint
at a fixed time on the Fock space:
\be
\int d{\bf q}\,\phi^{\dagger}(\sigma,{\bf q})\phi(\sigma,{\bf q})
-\rho(\sigma)=0,
\ee
where $\rho$ is the composite field
\be
\rho=\frac{1}{2} \bar{\psi}(1-\sigma_{3})\psi,
\ee
which is equal to one on solid lines and zero on the dotted lines.
This constraint ensures that there is at most one solid line at each site,
thereby avoiding  problem a).

To avoid the unwanted propagators of b), we define, for any two lines
located at $\sigma_{i}$ and $\sigma_{j}$, with $\sigma_{j}>\sigma_{i}$,
\be
\mathcal{E}(\sigma_{i},\sigma_{j})=\prod_{k=i+1}^{k=j-1}\left(1 -
\rho(\sigma_{k})\right).
\ee
If $\sigma_{j}<\sigma_{i}$, $\mathcal{E}$ is defined to be zero. The
following property of this function that will be needed:
 $\mathcal{E}(\sigma_{i},\sigma_{j})$ is equal to one
only if the two solid lines at $\sigma_{i}$ and $\sigma_{j}$ are
seperated only by dotted lines. If there is one or more solid lines
in between, it is zero. If we now redefine $H_{0}$ as
\bq
H_{0}&=&\frac{1}{2}\sum_{\sigma,\sigma'} \int d{\bf q} \int d{\bf q'}
\,\frac{\mathcal{E}(\sigma,\sigma')}{\sigma' -\sigma} \left(
({\bf q} -{\bf q'})^{2} + m^{2}\right)\nonumber\\
&\times&  \phi^{\dagger}(\sigma,
{\bf q})\, \phi(\sigma,{\bf q})\,  \phi^{\dagger}(\sigma',
{\bf q'})\, \phi(\sigma',{\bf q}).
\eq
By applying $H_{0}$ to a state with several solid lines, it is easy
to see that $\mathcal{E}(\sigma,\sigma')$ projects out all the 
unwanted propagators.

There remains the problem c), the problem of the missing prefactor.
As explained in reference [1], it is best to attach this factor to the
vertices. Consider two types of vertices, corresponding to the 
beginning and ending of a solid line, pictured in Fig.3.
\begin{figure}[t]
\centerline{\epsfig{file=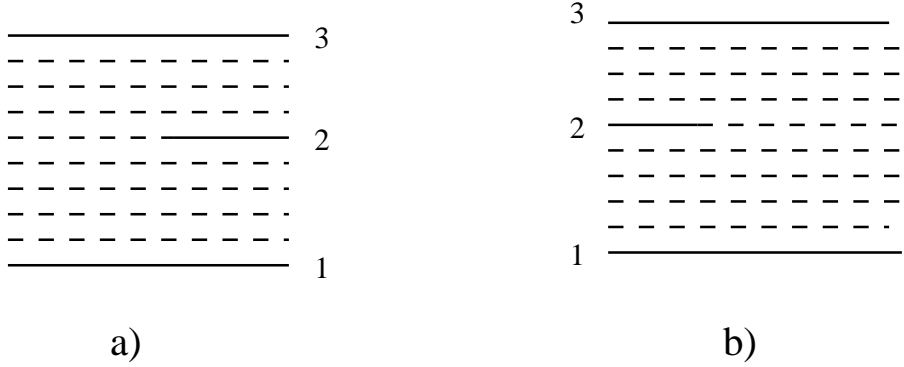, width=12cm}}
\caption{The Two $\phi^{3}$ Vertices}
\end{figure}
 The solid
lines are labeled as 1, 2 and 3, and the momenta that enter the 
vertices are labeled by the corresponding pair of indices 12, 23 and
13 respectively. Attaching a factor of
\be
V=\frac{1}{\sqrt{8\,p_{12}^{+} p_{23}^{+} p_{13}^{+}}}=
\frac{1}{\sqrt{8(\sigma_{2}-\sigma_{1}) (\sigma_{3}-\sigma_{2})
(\sigma_{3}-\sigma_{1})}}
\ee
to each vertex, takes care of the missing prefactors. However, we still face
a problem similar to the one encountered with the construction of $H_{0}$.
In the vertices of Fig.3 , the solid lines 1 and 3 at one end of the
vertex should be seperated by
only the dotted lines. To ensure this, we define
\be
\mathcal{V}(\sigma_{2})=\sum_{\sigma_{1}<\sigma_{2}}\sum_{\sigma_{2}<
\sigma_{3}}\frac{\rho(\sigma_{1}) \mathcal{E}(\sigma_{1},\sigma_{3})
\rho(\sigma_{3})}{\sqrt{8(\sigma_{2}-\sigma_{1}) (\sigma_{3}-\sigma_{2})
(\sigma_{3}-\sigma_{1})}}.
\ee
The numerator of this expression picks the correct vertex configuration, and
it projects out all the other unwanted configurations. With the help of
this vertex, we can rewrite the final form of $H_{I}$, which includes
the additional factor $V$ of eq.(10):
\be
H_{I}= g \sum_{\sigma} \int d{\bf q}\left(\mathcal{V}(\sigma)\,\phi(\sigma,
{\bf q})\,\rho_{+}(\sigma) + \rho_{-}(\sigma)\,\phi^{\dagger}(
\sigma,{\bf q})\,\mathcal{V}(\sigma)\right),
\ee
where,
$$
\rho_{\pm}=\frac{1}{2} \bar{\psi}(\sigma_{1}\pm i \sigma_{2}) \psi.
$$
The additional factors $\rho_{\pm}$ are needed to make sure that a
 solid line in
the Fock space is always paired with an $i=2$ fermion and a dotted line
with an $i=1$ fermion.

Now that the various pieces that make up the total Hamiltonian are in
place, we define,
\be
H= H_{0}+ H_{I} +H',
\ee
where $H_{0}$ and $H_{I}$ are given by eqs.(9) and (12), and,
\be
H'=\sum_{\sigma}\left(\int d{\bf q}\, \phi^{\dagger}(\sigma,{\bf q})\,
\phi(\sigma,{\bf q}) -\rho(\sigma)\right)\,\lambda(\sigma),
\ee
implements the constraints (6)  by means of a Lagrange multiplier
$\lambda$.

Taking advantage of (6), it is possible to rewrite the free
Hamiltonian in a somewhat simpler form:
\bq 
H_{0}&=&\frac{1}{2} \sum_{\sigma \neq \sigma'}
 G(\sigma,\sigma')\Big(\rho(\sigma)
\,\int d{\bf q'}\,{\bf q'}^{2}\, \phi^{\dagger}(\sigma',
{\bf q'})\, \phi(\sigma',{\bf q'})
+\frac{1}{2} m^{2}
\rho(\sigma) \rho(\sigma')\nonumber\\
&-& \int d{\bf q} \int d{\bf q'}\, ({\bf q}\cdot {\bf q'})\,
 \phi^{\dagger}(\sigma,
{\bf q})\, \phi(\sigma,{\bf q}) \, \phi^{\dagger}(\sigma',
{\bf q'})\, \phi(\sigma',{\bf q'})
\Big),
\eq
where, to simplify writing, we have defined,
$$
G(\sigma, \sigma')=\frac{\mathcal{E}(\sigma,\sigma')+\mathcal{E}(\sigma',
\sigma)}{|\sigma-\sigma'|}.
$$

Although we will mostly stick with the Hamiltonian picture in this paper,
if so desired, one can switch to the path integral approach based on
the action
\be
S=\int d\tau \left(\sum_{\sigma}\left(i\bar{\psi}\partial_{\tau}\psi
+i\int d{\bf q}\,\phi^{\dagger} \partial_{\tau}\phi\right)- H(\tau)
\right).
\ee

\vskip 9pt

\noindent{\bf 4. Phase Invariance And Bosonization}

\vskip 9pt

It is easy to verify that the above action is invariant under the following
phase transformation:
\bq
\psi &\rightarrow& \exp\left(-\frac{i}{2}\,\alpha\,\sigma_{3}\right)\,
\psi,\,\,\,\bar{\psi} \rightarrow \bar{\psi}\,\exp\left(\frac{i}{2}\,
\alpha\,\sigma_{3}\right),\nonumber\\
\phi&\rightarrow &\exp(-i\,\alpha\,)\,\phi,\,\,\,\phi^{\dagger}
\rightarrow \exp(i\,\alpha\,)\,\phi^{\dagger},\,\,\, \lambda\rightarrow
\lambda - \partial_{\tau}\alpha.
\eq 
Here $\alpha$ is an arbitrary function of $\sigma$ and $\tau$, so this
is a gauge transformation on the world sheet. By a suitable gauge fixing,
it should be possible to eliminate one of the degrees of freedom of
the fermionic field $\psi$. To see how this comes about, it is very
convenient first to bosonize $\psi$. In addition to $\rho$ (eq.(7)),
we introduce the bosonic field $\xi$ and set,
\bq
\bar{\psi}\sigma_{3}\psi&=& 1-2\,\rho,\,\,\,\bar{\psi}\sigma_{1}\psi=
2\,\sqrt{\rho -\rho^{2}}\,\cos(\xi),\nonumber\\
\bar{\psi}\sigma_{2}\psi&=& 2\,\sqrt{\rho -\rho^{2}}\,\sin(\xi).
\eq
The first equation is simply a rewrite of eq.(7). The kinetic energy
term for $\psi$ in eq.(16) can be replaced by its bosonic counterpart:
\be
\int d\tau \sum_{\sigma} i\,\bar{\psi}\partial_{\tau}\psi \rightarrow
\int d\tau \sum_{\sigma} \xi\,\partial_{\tau}\rho.
\ee
One can check that this action produces the correct equations of motion
and the correct commutation relations for the fermionic bilinears.

We note that bosonization has replaced discrete variables by continous
ones. For example, according to its original definition as a composite
field (eq.(7)), $\rho$ could only take on the values 0 and 1, but as an
independent bosonic field, it can vary continuously between 0 and 1.
It is natural to interpret it as the probability of finding a solid
line at a given location. As we shall see in the next section, the
reformulation of the problem in terms of continuous variables provides
a convenient setup for the mean field approximation.

Now consider the effect of the phase transformation (17) on the bosonic
fields: $\rho$ is unchanged, whereas $\xi$ transforms according to
$$
\xi \rightarrow \xi+\alpha,
$$
and therefore, we can gauge fix by setting
\be
\xi=0.
\ee
In this gauge, and with fermions
bosonized, the interaction Hamiltonian (12) becomes
\be
H_{I}\rightarrow g\,\sum_{\sigma}\,\mathcal{V}(\sigma)\,\sqrt{\rho(\sigma)
-\rho^{2}(\sigma)}\,\int d{\bf q}\,\left(\phi(\sigma,{\bf q})+
\phi^{\dagger}(\sigma,{\bf q})\right). 
\ee

Although the field $\xi$ disappeared from the problem, its equation of
motion, namely
\be
\partial_{\tau}\rho= 0,
\ee
has to be imposed as a constraint. This constraint means that being time
independent, $\rho$ is no longer a dynamical field. Recalling that we have
not specified the initial conditions, we can do so now by assigning an
arbitrary probability distribution $\rho(\sigma)$ for the solid lines at
some initial time. This probability distribution is then constant in time by
virtue of the above
constraint. In the next section, $\rho(\sigma)$ will be
determined in the mean field approximation by minimizing the ground state
 energy. 

\vskip 9pt

\noindent{\bf 5. The Mean Field Approximation}

\vskip 9pt

The Hamiltonian (13) is exact but quite complicated; for example, it is
non-local in the coordinate $\sigma$. Clearly, it is not a good starting
point for doing the usual Feynman perturbation expansion. The standard
perturbation theory is an expansion in the $\rho_{0}=0$ phase of the model.
Instead, we are here interested in the phase where Feynman graphs are dense
on the world sheet, with $\rho_{0}\neq 0$. In this new phase,  the
Hamiltonian (13), in conjunction with the mean field method, turns out to be
an excellent starting point for doing calculations. In fact,
in the standard approach, one would be at a loss to even define the new
phase precisely.

 In reference [1], a variational approach
was used to calculate the ground state of (13). Here, instead, 
 the mean field approximation  will enable us 
to compute, in addition to
the ground state,  the excited modes of the model. We should point out 
 right at the beginning that,
 for the meanfield method to make sense, we have to take the
total number of lines on the world sheet,
\be
N_{0}= p^{+}/a,
\ee
to be large but finite. So we are close to the
world sheet continuum limit, but 
$a$ is always kept non-zero to avoid ill defined expressions.

The mean field approximation amounts to replacing every bosonic field
in the Hamiltonian  by its  ground state expectation value, and then
minimizing the resulting ground state energy. The subscript ``0'' will
 indicate the expectation value of the corresponding field; for example,
 $\langle \rho \rangle =\rho_{0},\,\langle \phi \rangle= \phi_{0}$, and so on.
We will assume that the ground state is invariant under translations of
$\sigma$ and $\tau$, so that the expectation values of the fields do not
depend on these variables. Therefore, $\rho_{0}$ and $\lambda_{0}$ 
 are constants, whereas $\phi_{0}$ and
 $\phi^{\dagger}_{0}$ depend only on ${\bf q}$. Furthermore, by rotation 
invariance, they can only depend on the length of the vector ${\bf q}$.
We also note that in view of the discussion at the end of the last section,
$\rho$ is not a dynamical field but it merely serves to fix the initial
conditions. Letting $\rho \rightarrow \rho_{0}$ is an exact replacement;
there are no fluctuations around $\rho_{0}$. This means that we have once
for all fixed the initial conditions so as to minimize the ground state
energy. 

Replacing every bosonic field by its expectation value, as indicated
above, considerably simplifies the Hamiltonian. Let us start with the
free Hamiltonian $H_{0}$. Setting
$\rho=\rho_{0}$ in the definition of $\mathcal{E}$, we have,
\bq
\mathcal{E}(\sigma,\sigma')&=&(1-\rho_{0})^{n},\,\,\,n=(\sigma'
-\sigma)/a -1,\nonumber\\
G(\sigma,\sigma')&=&\frac{(1-\rho_{0})^{n}}{a (n+1)},\,\,\,\sigma'>\sigma,
\nonumber\\
\sum_{\sigma'>\sigma} G(\sigma,\sigma')&=&
-\frac{\ln(\rho_{0})}{a (1-\rho_{0})}.
\eq
In writing this equation, we have assumed that the sum over $n$
extends all the way to infinity, whereas in reality, there is an upper
cutoff of the order of $N_{0}$ (eq.(23)). But since $N_{0}$ is very large,
this  makes a difference only for very small values of $\rho_{0}$. In what
 follows, we will always keep $\rho_{0}$ away from zero. Substituting the
above result in $H_{0}$, we get,
\be
E_{0}=\langle H_{0} \rangle= N_{0} F(\rho_{0})\left(\int d{\bf q}\,
{\bf q}^{2}\, |\phi_{0}({\bf q})|^{2} +\frac{1}{2} m^{2}\,\rho_{0}\right),
\ee
where we have defined,
$$
F(\rho_{0})=-\frac{\rho_{0}\,\ln(\rho_{0})}{a (1-\rho_{0})}.
$$
Notice that the last term on the right hand side of eq.(15) vanishes because
of the rotation invariance of $\phi_{0}({\bf q})$.

Next, we focus on $H_{I}$ (eq.(21)), replacing $\rho$ by $\rho_{0}$ and 
$\phi$ by $\phi_{0}$.ith this replacement,
 $\mathcal{V}$  becomes
independent of $\sigma_{2}$; it depends implicitly only on
$\rho_{0}$:
\bq
\mathcal{V}&=& \frac{\rho_{0}^{2}}{\sqrt{8 a^{3}}}\,W(\rho_{0}),
\nonumber\\
W(\rho_{0})&=&\sum_{n_{1}=0}^{\infty}\sum_{n_{2}=0}^{\infty} \frac{
(1-\rho_{0})^{n_{1}+n_{2}+1}}{\sqrt{(n_{1}+1) (n_{2}+1) (n_{1}+ n_{2}+2)}}.
\eq
 The ground state  value of $H_{I}$, $E_{I}$, is then
given by
\be
E_{I}= N_{0}\,g\,\mathcal{V}(\rho_{0})\,\sqrt{\rho_{0}-\rho_{0}^{2}}\,
\int d{\bf q}\,\left(\phi_{0}({\bf q})+\phi_{0}^{*}({\bf q})\right),
\ee
and the total ground state energy $E_{t}$ by
\be
E_{t}= E_{0}+ E_{I} + N_{0}\,\lambda_{0}\,\left(\int d{\bf q}\,
|\phi_{0}({\bf q})|^{2} - \rho_{0}\right),
\ee
where $E_{0}$ is given by (25).

The next step is to write down the classical equations of motion 
that result from varying $E_{t}$ with respect to 
the expectation values of the fields.
The solution to these equations will then determine the ground
state energy. We will first write down the equations gotten by
varying $E_{t}$ with respect to $\phi_{0}$ and $\lambda_{0}$: 
 
\bq
\frac{\partial E_{t}}{\partial \phi_{0}({\bf q})}=0 &\rightarrow&
\phi_{0}=\phi_{0}^{*}=
-g\,\frac{\sqrt{\rho_{0}-\rho_{0}^{2}}\,\mathcal{V}(\rho_{0})}
{\lambda_{0}+F(\rho_{0})\,{\bf q}^{2}},\\
\frac{\partial E_{t}}{\partial \lambda_{0}}=0 &\rightarrow&
\rho_{0}=g^{2}\,\mathcal{V}^{2}(\rho_{0})\,(\rho_{0}-\rho_{0}^{2})\,
\int d{\bf q}\,\left(\lambda_{0}+F(\rho_{0})\,{\bf q}^{2}\right)^{-2}.
\eq
Apart from some redefinition of constants, $\phi_{0}({\bf q})$
is the same as the variational wave function $A({\bf q})$ of reference
[1]. This  suggests that  the mean field and the 
variational approximations are closely related.

Our goal is to eliminate all subsidiary variables except for $\rho_{0}$,
and to express $E_{t}$
in terms of only this remaining variable. We will then search for the
minimum value of $E_{t}$ and see whether this is realized for a
 value of  $\rho_{0}$ different from zero and one.
 As explained earlier, if $\rho_{0}$
turns out to be zero, we then have a trivial ground state corresponding
to an empty world sheet.  $\rho_{0}=1$ corresponds to a world sheet
where every line is an eternal solid line, which is simply a bunch of
free propagators and therefore trivial.
 In contrast, a  value for $\rho_{0}$ different from zero or unity
means that the ground state corresponds to a world sheet densely
covered by interacting Feynman graphs.
 This is the kind of ground state that can 
lead to interesting phenomena, such as string formation.
 
  Equation (30) involves an integral over ${\bf q}$, and 
 in order to proceed further,   the dimension
 $D$ of the transverse space, which has
 been arbitrary up to now, has to be specified. We will investigate this 
equation for $D=2$ and $D=4$ in the next section.

\vskip 9pt
\noindent{\bf 6. The Ground State At $D=2$ And $D=4$}
\vskip 9pt

$D=2$ (3+1 space-time dimensions) and $D=4$ (5+1 space-time dimensions) are
two interesting choices for the $\phi^{3}$ theory. In the first case, the
model is superrenormalizable; the coupling constant is finite, and there is
only a logarithmic mass divergence.  Eqs. (29) and (30)
can  be used to eliminate $\lambda_{0}$ and $\phi_{0}$, 
 and therefore the ground state energy 
$E_{t}$ can  be expressed solely in terms of $\rho_{0}$. In contrast,
at $D=4$, the model is renormalizable and asymptotically free. In this case,
$\lambda_{0}$ stays in the problem and it can be used to define
the renormalized coupling constant through eq.(38).
 This is a familiar situation, related to asymptotic freedom and the
running of the coupling constant. All of this is in complete agreement
with the well known results from perturbation theory.

Let us now set $D=2$ in eq.(30). The 
integral is finite, and after doing it, the result can be written as
\be
\lambda_{0}=\pi\,g^{2}\,\mathcal{V}^{2}(\rho_{0})\,\frac{1-\rho_{0}}
{F(\rho_{0})}.
\ee
 Combining this with (29), $E_{t}$ can be expressed as
a function of only $\rho_{0}$. The result, however, contains two integrals
$$
\int d{\bf q}\,\phi_{0}({\bf q}),\,\,\,\int d{\bf q}\,{\bf q}^{2}\,
\phi_{0}^{2}({\bf q}),
$$
which diverge logarithmically for large ${\bf q}$.
We evaluate these using a cutoff $\Lambda$ in $|{\bf q}|$, and obtain the
following final result for the ground state energy:
\be
E_{t}=N_{0}\left(-\pi\,g^{2}\,\mathcal{V}^{2}(\rho_{0})\,\frac{\rho_{0}
-\rho_{0}^{2}}{F(\rho_{0})} \left(\ln\left(\frac{F\,\Lambda^{2}}
{\lambda_{0}}\right)+1\right)+\frac{1}{2}\,m^{2}\,\rho_{0}\,
F(\rho_{0})\right).
\ee 

 To minimize the energy,  we search for the solutions of
\be
\frac{\partial E_{t}}{\partial \rho_{0}}=0.
\ee
We note that, in the limit $\Lambda\rightarrow \infty$,
 the location of the minimum is solely determined by the cutoff dependent
term in $E_{t}$.
At the same time, we would like to introduce a mass counter term to
cancel the cutoff dependence. This is done by replacing 
  $m$ in (25) by the bare mass $m_{0}$ and setting
\be
m_{0}^{2}= s\,\ln\left(\Lambda^{2}/\mu^{2}\right),
\ee
where $s$ is a parameter which will later be ajusted to cancel the divergence.
One has first to minimize the cutoff dependent terms in
 $E_{t}$ at fixed $s$ and $\Lambda$ with respect
to $\rho_{0}$, and then adjust $s$ so that in the final expression, the
logarithmic divergence cancels. This leads to two simultaneous equations
for the two parameters $\rho_{0}$ and $s$:
\bq
&&-\pi\,g^{2}\,\frac{\partial}{\partial \rho_{0}}\left(\mathcal{V}^{2}
(\rho_{0})\,\frac{\rho_{0} -\rho_{0}^{2}}{F(\rho_{0})}\right)+
\frac{1}{2}\,s\,\frac{\partial}{\partial \rho_{0}}\left(\rho_{0}\,
F(\rho_{0})\right)=0,\nonumber\\
&&-\pi\,g^{2}\,\mathcal{V}^{2}(\rho_{0})\,
\frac{(\rho_{0}-\rho_{0}^{2})}{F(\rho_{0})}+\frac{1}{2}\,s\,
\rho_{0}\,F(\rho_{0})=0.
\eq
 Eliminating $s$,  the two equations reduce
to a single equation:
\be
\frac{\partial L(\rho_{0})}{\partial \rho_{0}}=0,
\ee
where $L$ is given by
$$
L(\rho_{0})=\frac{W^{2}(\rho_{0})\,
\rho_{0}^{2}\,(1-\rho_{0})^{3}}{\ln^{2}(\rho_{0})},
$$
and $W$  by (26). Now,
$$
W(\rho_{0})\rightarrow \pi^{3/2}\,(\rho_{0})^{-1/2}
$$
as $\rho_{0}\rightarrow 0$, and
$$
W(\rho_{0})\rightarrow 2^{-1/2}\,(1-\rho_{0})
$$
as $\rho_{0}\rightarrow 1$. The function $L$,
schematically plotted as a function of $x=\rho_{0}$
 in Fig.(4),
\begin{figure}[t]
\centerline{\epsfig{file=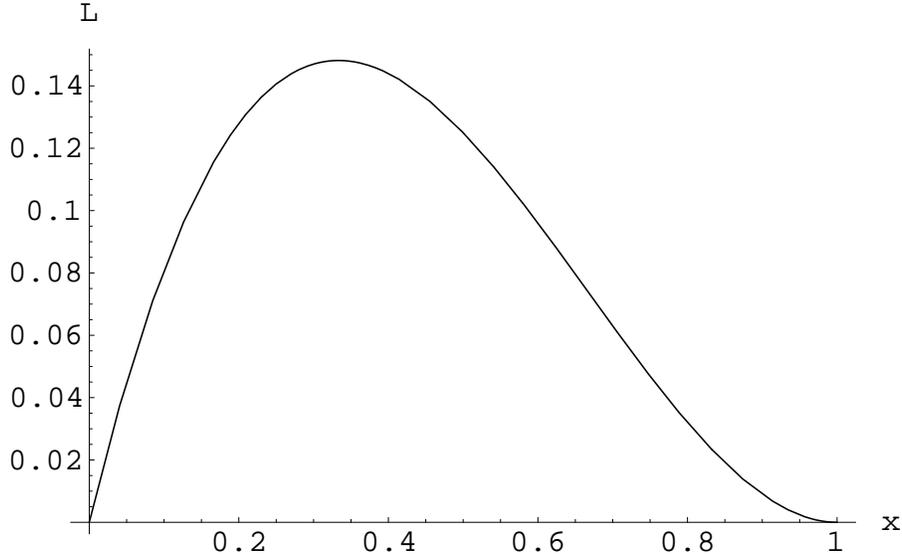, width=12cm}}
\caption{The Function L(x)}
\end{figure}
 vanishes at both
$\rho_{0}= 0$ and $\rho_{0}= 1$, and it is positive in between. Its
derivative has a single zero in the interval, corresponding to cutoff
independent minimum for $E_{t}$ at $\rho_{0}\neq 0$. The coefficient
of the logarithmic cutoff term in $E_{t}$, 
\be
Z(\rho_{0})=
-\pi\,g^{2}\,\mathcal{V}^{2}(\rho_{0})\,\frac{\rho_{0}
-\rho_{0}^{2}}{F(\rho_{0})}+\frac{1}{2}\,s\,\rho_{0}\,
F(\rho_{0}),
\ee
is plotted schematically
 as a function of $x=\rho_{0}$ in Fig.(5).
\begin{figure}[t]
\centerline{\epsfig{file=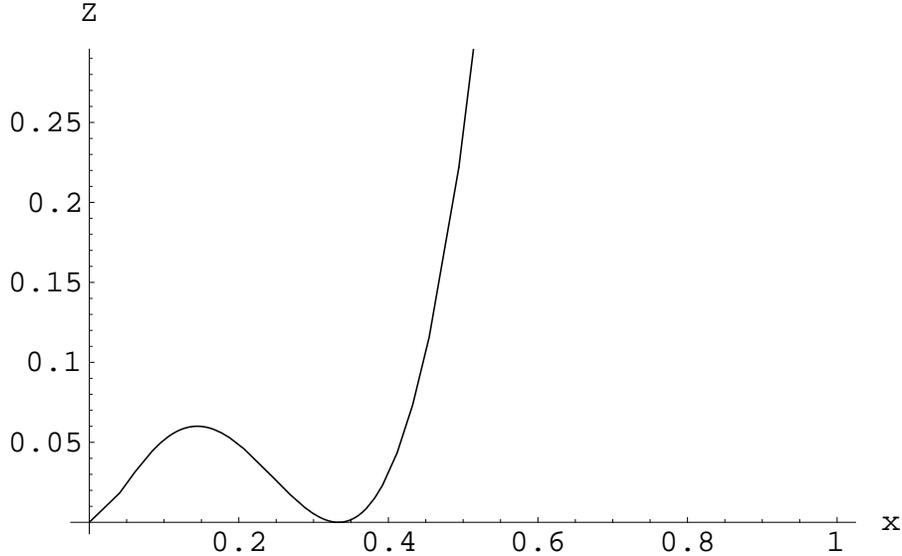, width=12cm}}
\caption{The Function Z(x)}
\end{figure}
 There are 
two minima, a trivial one at $\rho_{0}=0$, and the non-trivial one
at some point in the interval $0<\rho_{0}<1$. Although these minima appear
 degenerate, that is because $Z$ does not include
 the cutoff independent terms in $E_{t}$. These terms have no influence
on the position of the minima, but the value of $E_{t}$ at the
non-trivial minimum does depend on them. For example, we can change it
 by adding an arbitrary cutoff independent term to $m_{0}^{2}$ in (30).
In this way, we can adjust $E_{t}$ to be negative at the non-trivial
minimum, so that it becomes the true global minimum.
We conclude that  one can arrange to have a
global minimum of the energy which is cutoff independent
 at a non-zero value of $\rho_{0}$. 

Let us now consider the case $D=4$. The integral that appears in 
 equation  (30) is now logarithmically divergent. We introduce a cutoff
$\Lambda$ as before, and after doing the integral, the result can be
written in the form,
\be
\frac{1}{g_{0}^{2}}=\pi^{2}\,\mathcal{V}^{2}(\rho_{0})\,\frac{1-\rho_{0}}
{F^{2}(\rho_{0})}\,\left(\ln\left(\frac{\rho_{0}\,F\,\Lambda^{2}}{
\lambda_{0}}\right)-1\right),
\ee
where $g$ has been replaced by $g_{0}$ in anticipation of 
renormalization. It is now natural to define the renormalized coupling
constant $g_{r}$ at the energy scale corresponding to $\lambda_{0}$:
by
\be
g_{r}^{2}= g_{0}^{2}\,\ln\left(\frac{\Lambda^{2}}{\lambda_{0}}\right).
\ee
This equation is consistent with the well known asymptotic freedom
of the $\phi^{3}$ theory in $5+1$ dimensions. Also, the results derived here,
using the mean field method, in particular the expression for
 $\phi_{0}$ (eq.(29)), are in agreement with those derived
in [1] using a variational calculation.

Just as in the case $D=2$,
the ground state energy $E_{t}$ can be calculated by eliminating the
auxilliary variables. 
 The integrals over ${\bf q}$ that one encounters
 are now quadratically divergent, and the leading cutoff dependent piece
plus the mass term in $E_{t}$ is given by
\be
E_{t}\approx N_{0}\,\left(-\pi^{2}\,g^{2}\,\mathcal{V}^{2}(\rho_{0})\,\frac{
\rho_{0}-\rho_{0}^{2}}{F(\rho_{0})}\,\Lambda^{2}+ \frac{1}{2}\, m_{0}^{2}\,
\rho_{0}\,F(\rho_{0})\right).
\ee
Apart from a different power of $\pi$, the coefficient of $\Lambda^{2}$
is identical to the coefficient of $\ln(\Lambda)$ in (32) for $D=2$.
Consequently, letting
$$
m_{0}^{2}=s\,\Lambda^{2},
$$
we end up with   eq.  (36),
 and the same minimum $\rho_{0}\neq 0$ as before.

\vskip 9pt

\noindent{\bf 7. Fluctuations Around The Mean Field}

\vskip 9pt

In the last section, using the mean field approximation,
 we found a non-trivial ground
state corresponding to non-vanishing bosonic fields. The next step
would be to treat this solution as a classical background, and to calculate
the quadratic fluctuations about this background. This would then give us
the spectrum of the free theory based on this new ground state. The cubic
and higher order terms in the fluctuations will be responsible for the
interactions and they will not be considered here. We will
see in this section that, the spectrum of the fluctuating modes consists
of two sectors: A continuum and discrete bound states. It is the bound states
that generate the string spectrum, which eventually merges into the
continuum. In an appropriately defined  limit
$N_{0}\rightarrow \infty$, the continuum can be pushed
all the way up, leaving behind the spectrum
of a free bosonic string. As discussed in the introduction, this is not
unexpected: When the world sheet is uniformly and
 densely covered by graphs, it is reasonable to expect that it can be
 described by an effective Nambu action, leading to a string picture.

In what follows, we will fix 
 the fields $\lambda$ and
$\rho$ at their classical values $\lambda_{0}$ and
$\rho_{0}$ respectively. For the field $\rho$, this is no restriction; as
we have argued earlier, $\rho$ is fixed by the boundary conditions and does
not fluctuate. $\lambda$ could in principle fluctuate; however, since this
field carries no momentum, it is  not expected to contribute to the string
trajectories. The only fields that carry momentum are $\phi(\sigma,{\bf q})$
and $\phi^{\dagger}(\sigma,{\bf q})$, and we will calculate to second order
 the fluctuations of these fields around their
ground state expectation values.

Defining $H_{p}$ the collection of $\phi$ dependent terms in the
Hamiltonian, we have,
\bq
H_{p}&=&\sum_{\sigma,\sigma'}\,\frac{1}{2}\,
 G(\sigma,\sigma')\,\Big(\rho_{0}\,\int d{\bf q}\,{\bf q}^{2}\,
 \phi^{\dagger}(\sigma,{\bf q})\, \phi(\sigma,{\bf q})+
\lambda_{0}\,\int d{\bf q}\, \phi^{\dagger}(\sigma,{\bf q})\,
 \phi(\sigma,{\bf q})\nonumber\\
&-&\int d{\bf q}\,\int d{\bf q'}\,({\bf q}\cdot{\bf q'})
 \phi^{\dagger}(\sigma,
{\bf q})\, \phi(\sigma,{\bf q}) \, \phi^{\dagger}(\sigma',
{\bf q'})\, \phi(\sigma',{\bf q'})\Big).
\eq
We have not included $H_{I}$ since it is linear in $\phi$ and $\phi^{\dagger}$
and consequently does not contribute to the quadratic terms in fluctuations.
 Before expanding $H_{p}$ around the classical background, it
is convenient to define
$$
\phi_{r}=\frac{1}{2}(\phi+\phi^{\dagger}),\,\,\,
\phi_{i}=\frac{1}{2 i}(\phi-\phi^{\dagger}),
$$
and,
\bq
\phi_{r}(\sigma,{\bf q})&=&\phi_{0}({\bf q})+\left(
\frac{1}{2}\,\left(
F(\rho_{0})\,{\bf q}^{2}+\lambda_{0}\right)\right)^{1/2}
\,\chi_{1}(\sigma,{\bf q}),\nonumber\\
\phi_{i}(\sigma,{\bf q})&=&\left(
2\,\left(F(\rho_{0})
\,{\bf q}^{2}+\lambda_{0}\right)\right)^{- 1/2}
\,\chi_{2}(\sigma,{\bf q}),
\eq
where $\chi_{1,2}$ are the fluctuating fields and
the background  $\phi_{0}$ is given by (29).

The terms quadratic in $\chi_{1,2}$ in the Hamiltonian are,
\be
H^{(2)}=\frac{1}{2}\,\sum_{\sigma}\int d{\bf q}\,
\chi_{2}^{2}(\sigma,{\bf q})
+\sum_{\sigma,\sigma'} \int d{\bf q} \int d{\bf q'}\,
\chi_{1}(\sigma,{\bf q})\,M(\sigma,{\bf q},\sigma',{\bf q'})\,
\chi_{1}(\sigma',{\bf q'}),
\ee
where $G$ is given by (18) and $M$ by
\bq
&M(\sigma,{\bf q}, \sigma', {\bf q'})&
=\frac{1}{2}\,\left(F\,{\bf q}^{2}+\lambda_{0}\right)^{2}
\,\delta_{\sigma,\sigma'}\,\delta({\bf q}-{\bf q'})\nonumber\\
&-\, G(\sigma,\sigma')& 
\left(F\,{\bf q}^{2}+\lambda_{0}\right)^{1/2}\,\phi_{0}({\bf q})\,
({\bf q}\cdot {\bf q'})\,\left(F\,{\bf q'}^{2}+\lambda_{0}\right)^{1/2}\,
\phi_{0}({\bf q'}).
\eq

The energy levels of the excited states are determined by diagonalizing
$M$. The eigenfunctions can be written as a product:
$$
\chi_{1}\rightarrow f_{\eta}(\sigma)\,h_{\omega,\eta}({\bf q}),
$$
where $f$ is an eigenstate of $G$
\be
\sum_{\sigma'} G(\sigma,\sigma')\,f_{\eta}(\sigma')=\eta\,f_{\eta}(\sigma')
\ee
with eigenvalue $\eta$, and $h$ satisfies,
\bq
\omega\, h_{\omega,\eta}({\bf q})&=&
\left(F\,{\bf q}^{2}+\lambda_{0}\right)^{2}
\, h_{\omega,\eta}({\bf q})
- 2 \eta\, \left(F\,{\bf q}^{2}+\lambda_{0}\right)^{1/2}
\,\phi_{0}({\bf q})\nonumber\\
&\times&
\int d{\bf q'}\,({\bf q}\cdot {\bf q'})\,\left(F\,{\bf q'}^{2}+\lambda_{0}
\right)^{1/2}\,
\phi_{0}({\bf q'})\, h_{\omega,\eta}({\bf q'}).
\eq

The next step is to solve eqs. (45) and (46). We first focus on (46). The
solutions fall into two classes:\\
a) $\omega$ is in the continuum. The solution is given by
\be
 h_{\omega,\eta}({\bf q})=\delta({\bf q}^{2}-\beta^{2})\,
\tilde{h}_{\omega}({\bf q}),
\ee
where $\beta$ is a real number, and
the function $\tilde{h}_{\omega}$ has to satisfy
\be
\int d{\bf q}\,\delta({\bf q}^{2}-\beta^{2})\,{\bf q}\,\tilde{h}_{\omega}
({\bf q})=0,
\ee
but is otherwise arbitrary. $\omega$ is given by
\be
\omega=\left(F(\rho_{0})\,\beta^{2}+\lambda_{0}\right)^{2}.
\ee
Keeping in mind that both $\lambda_{0}$ and $F$ are positive, as $\beta$
goes from $0$ to $\infty$, $\omega$ varies continuously between
$\lambda_{0}\geq 0$ and $\infty$. This kind of continuous spectrum is
 expected from the perturbative analysis of the underlying
field theory.\\
b) Now suppose the integral in (48) does not vanish. In this case, the
solution to eq.(46) can be written as
\be
h^{i}_{\omega}({\bf q})=2\,\eta\,C\,q^{i}\,\frac{\left(F\,{\bf q}^{2}+
\lambda_{0}\right)^{1/2}\,\phi_{0}({\bf q})}{
\left(F\,{\bf q}^{2}+\lambda_{0}\right)^{2}\,-\omega}.
\ee
$C$ is an arbitrary normalization constant and
 the index $i$ ranges from $1$ to $D$ over the transverse space, giving rise
to $D$ independent solutions. Substituting this back in (46) gives
the consistency condition
\be
q^{i}=2\,\eta\,\int d{\bf q'}\,{\bf q'}^{i}\,({\bf q}\cdot {\bf q'})\,
\frac{\left(F\,{\bf q'}^{2}+\lambda_{0}\right)\,\phi_{0}^{2}({\bf q'})}
{\left(F\,{\bf q'}^{2}+\lambda_{0}\right)^{2}\,-\omega}.
\ee

So far, $D$ has been arbitrary, but now we specialize to the case
$D=2$. The consistency condition then reads
\be
1=\frac{\pi\,\eta\,g^{2}\,(\rho_{0}-\rho_{0}^{2})\,\mathcal{V}^{2}}
{\lambda_{0}\, F^{2}}\,K(\bar{\omega}),
\ee
where $\bar{\omega}=\omega/\lambda_{0}^{2}$ and
\be
K(\bar{\omega})=\int_{0}^{\infty} d x\,\frac{x}{(x+1)\,\left((x+1)^{2}
-\bar{\omega}\right)}.
\ee
The integral can be evaluated but it is just as easy to deal with it
directly. Substituting the expression for $\lambda_{0}$ given by (31),
the consistency condition can be rewritten as
\be
1=\frac{\eta\,\rho_{0}}{F(\rho_{0})}\,K(\bar{\omega}).
\ee

The above equation, which will in general have discrete solutions,
 can be thought of as an eigenvalue equation for bound states.
We now study it for various ranges of values of 
$\bar{\omega}$. To do this, we have to know something about $\eta$.
We shall shortly show that
\be
0<\frac{\eta\,\rho_{0}}{F(\rho_{0})} \leq 2.
\ee
For the moment, let us assume this result and first
  consider the range $\bar{\omega} >1$. The integral (53) is then
ill defined, and if one uses the  $i\epsilon$ prescription to define
it, it will become complex. This is easy to understand; 
the dicussion following eq.(49) shows that the continuum starts
at  $\bar{\omega} =1$, and therefore
 the bound state is now sitting on top of the continuum and
can decay into it. The imaginary part of the integral (53) is the 
reflection of this instability.

 Next consider the range $0\leq \bar{\omega} \leq 1$. In this range,
the integral is well defined, and
$K$ varies monotonically from $K(0)=1/2$ to $K(1)=\ln(2)\approx 0.69$.
Then, for a suitable range of $\eta$ in the interval allowed by (55),
there is a unique solution for $\bar{\omega}$ for a given $\eta$.
For the rest of the paper, since we will only be interested  in the stable
bound states, this range for  $\bar{\omega}$ will be the focus of our
attention. In particular, there is a special solution with  $\bar{\omega}=0$,
which will play an important role in the subsequent development.
 We will  shortly see that there exists an 
$\eta$ for which  
\be
\frac{\eta\,\rho_{0}}{F(\rho_{0})}=2,
\ee
and, for this $\eta$, recalling that $K(0)=1/2$,
the corresponding $\bar{\omega}=0$.

There is, of course, a special significance to a zero frequency oscillation
around a fixed background. In the case of classical solutions such as
solitons or instantons, the existence
of a zero mode is usually a consequence of a
symmetry, such as, for example, translation invariance, which is broken by
the classical solution. Integration over the zero mode restores this
symmetry. In the present case, the symmetry is translation invariance in
${\bf q}$: The Hamiltonian of eq.(13) is invariant under
\be
{\bf q}\rightarrow {\bf q}+{\bf r},
\ee
where ${\bf r}$ is a constant vector, whereas the classical solution
(29) for $\phi_{0}({\bf q})$ clearly breaks this symmetry. The
  $\bar{\omega}=0$ solution is then the Goldstone mode of this broken
symmetry. It is easy to show that, up to normalization,
 the corresponding $\chi_{1}$ is 
generated by an infinitesimal translation in ${\bf q}$ of 
 the background $\phi_{0}({\bf q})$:
$$
\chi_{1}^{i}\rightarrow \frac{\partial}{\partial q^{i}} \phi_{0}
({\bf q}). 
$$

 Finally, we have to consider the remaining range $\bar{\omega}<0$.
The existence of a solution for this range would be a disaster, since it
would correspond to a completely tachyonic spectrum. Fortunately, there is
no such solution. To show this, we observe that the integral (53) is a
monotonically increasing function of $\bar{\omega}$ for $\bar{\omega}<0$, so
that
$$
K(\bar{\omega})< K(0)=1/2
$$
for negative  $\bar{\omega}$. In view of the upper bound on $\eta$ given
by (55), the eigenvalue equation cannot be satisfied for negative 
 $\bar{\omega}$.

Most of the preceding development goes through in the case $D=4$, except
 the consistency condition (51) now involves a
logarithmically divergent integral:
\be
\frac{1}{g_{0}^{2}}=
\pi^{2}\,\eta\,\mathcal{V}^{2}\,\frac{\rho_{0}-\rho_{0}^{2}}
{2\,F^{3}}\,\int_{0}^{F\,\Lambda^{2}/\lambda_{0}} dx\,\frac{x^{2}}
{(x+1)\,\left((x+1)^{2}-\bar{\omega}\right)}.
\ee
We again search for the Goldstone mode by setting $\bar{\omega}=0$ in
the above integral and fixing $\eta$ by eq.(56). The result is
\be
\frac{1}{g_{0}^{2}}=\pi^{2}\,\mathcal{V}^{2}(\rho_{0})\,\frac{1-\rho_{0}}
{F^{2}(\rho_{0})}\,\left(\ln\left(\frac{F\,\Lambda^{2}}{
\lambda_{0}}\right)-\frac{3}{2}\right).
\ee
This agrees with eq.(38) for the running coupling constant except for
the constant term following the logarithm. This discrepency is due to
the naive cutoff we are using in regulating the integrals in eqs.(30)
 and (51).
This cutoff violates the translation invariance  responsible for the
zero mode. What is important is that the cutoff dependent
 logarithmic terms in (38) and (59) agree, and these are the only terms
that we will need later on. Using a
more refined scheme of regulation, it should also be possible to bring  the
constant terms into agreement. We have not tried to construct such a
 scheme, since these  will play no role in the subsequent development.

\vskip 9pt

\noindent{\bf 8. String Formation}

\vskip 9pt

So far, we have only found  zero mode and continuum
 solutions the basic eq.(46).
To find other bound states of interest, we will now proceed to solve
the eigenvalue equation (45). This will also enable us to verify the
statements (55) and (56). We make the ansatz
\be
f_{\eta}(\sigma)= e^{i k \sigma},
\ee
where k is a real number between $-\pi/a$ and $\pi/a$. The left hand side
of the eigenvalue equation can then be evaluated:
\bq
\sum_{\sigma'} G(\sigma,\sigma')\,e^{i k \sigma'}&=&\sum_{n=0}^{\infty}
\frac{(1-\rho_{0})^{n}}{a\,(n+1)}\,\exp\left(i\,k\,(\sigma+(n+1) a)\right)
+(k\leftrightarrow -k)\nonumber\\
&=&-\frac{e^{i k\sigma}}{a\,(1-\rho_{0})}\,\ln\left(\rho_{0}^{2}+
2\,(1-\rho_{0})\,(1-\cos(k\,a))\right),
\eq
from which it follows that,
\be
\eta(k)=-\frac{1}{a\,(1-\rho_{0})}\,\ln\left(\rho_{0}^{2}+
2\,(1-\rho_{0})\,(1-\cos(k\,a))\right),
\ee
and the ratio that in eq.(54) is given by
\be
\frac{\eta\,\rho_{0}}{F(\rho_{0})}=\frac{\ln\left(\rho_{0}^{2}+
2\,(1-\rho_{0})\,(1-\cos(k\,a))\right)}{\ln(\rho_{0})}.
\ee
From this expression, it is easy to verify  (55); and also, setting $k=0$,
eq.(56) follows.

As we have already stressed, we are mainly interested in the limit
of large $N_{0}$, which means small $a$, and we therefore 
look for solutions in the limit $ka\rightarrow 0$. In this limit,
$\eta\,\rho_{0}/F$ tends to 2, and $\bar{\omega}$ tends to zero.
Expanding the right hand side of (63) in powers of $k a$ gives
\be
\frac{\eta\,\rho_{0}}{F(\rho_{0})}\rightarrow
 2+\frac{1-\rho_{0}}{\rho_{0}^{2}\,\ln(\rho_{0})}\,(k\,a)^{2}.
\ee
We shall see that in the limit we are considering, higher order terms
in $k a$ will not contribute.

Up to this point, we have been studying the eigenfunctions and eigenvalues
of $G$, which do not depend on $D$. Now we turn our attention to $K$,
which does depend on $D$. As before, we will first take $D=2$.
 Since, in the limit $k a\rightarrow 0$, $\bar{\omega}$ tends to zero,
we can expand eq.(53) for
 $K$  to first order in $\bar{\omega}$:
\be
K(\bar{\omega})\rightarrow
\frac{1}{2}+ \frac{\bar{\omega}}{12},
\ee
and putting these results in eq.(54), we have,
\be
\omega \rightarrow - \frac{3\,(1-\rho_{0})\,\lambda_{0}^{2}}{\rho_{0}^{2}\,
\ln(\rho_{0})}\,(k\,a)^{2}
= \frac{3\,\pi^{2}\,g^{4}\,\rho_{0}^{4}\,(1-\rho_{0})^{5}\,
W^{4}(\rho_{0})}{64\,a^{2}\,|\ln(\rho_{0})|^{3}}\, k^{2}.
\ee
As we shall see shortly, the slope of the expected string depends on $\omega$.
If we insist on a finite slope in the limit of large $N_0$, small $a$,
 $\omega$ must remain finite in this limit. This requires the tuning
of $g$ by setting,
\be
g^{2}=a\,\bar{g}^{2}
\ee
and keeping $\bar{g}$ fixed and finite as $a\rightarrow 0$. This should
not be a surprise;  $a\rightarrow 0$ is the continuum limit on the world
sheet, and it is well known that, to get a sensible continuum limit,
 the parameters of a theory defined on a lattice have in general to be
fine tuned.

The tuning of the coupling constant by (67) has another desirable consequence.
From eq.(31), it follows that, as  $a\rightarrow 0$,
\be
\lambda_{0}\sim 1/a\sim N_{0},
\ee
and the continous spectrum which starts at
$$
\bar{\omega}=1,\,\,\,\omega=\lambda_{0}^{2}
$$
is pushed up to infinity. Therefore, the bound state energies (eigenvalues)
stay finite as the continuum threshold tends to infinity. Of course, for 
highly excited states with $k$ of the order of $1/a$, this argument breaks
down, but the energy of such states can be pushed up arbitrarily. We shall
 see that, in this limit, we shall have a string spectrum cleanly seperated
 from the high lying continuum.

For ease of exposition we have been treating $k$ as a continuous variable.
In reality since
 $\sigma$ is compactified on a circle of circumference $p^{+}$,
and $k$ is quantized according to
\be
k\,a= 2\,\pi\,n/N_{0},
\ee
where $n$ is an integer with $|n|\leq N_{0}$. We can therefore replace (60)
by the normalized
\be
f_{\eta}(\sigma)=\frac{1}{\sqrt{N_{0}}}\,e^{i\,k\,\sigma},
\ee
and choose $C$ in (50) so that the integral of the square of $h^{i}_{\omega}
({\bf q})$ is also normalized to unity. Let us now define the operators
\bq
Q^{i}_{k}&=&\sum_{\sigma}\int d{\bf q}\,\chi_{1}(\sigma,{\bf q})\,
h^{i}_{\omega}({\bf q})\,f_{\eta}(\sigma)\nonumber\\
P^{i}_{k}&=&\sum_{\sigma} \int d{\bf q}\,\chi_{2}(\sigma,{\bf q})\,
h^{i}_{\omega}({\bf q})\,f_{\eta}(\sigma).
\eq
We recall that both $\omega$ and $\eta$ are functions of $k$ through eqs.(66)
and (63), and $i$ labels the components of the vector in the transverse space.
$P$ and $Q$ satisfy the commutation relations
$$
[Q^{i}_{k},P^{i'}_{k'}]=i\,\delta_{k,k'}\,\delta_{i,i'}.
$$
These operators can be thought of as coordinates and momenta of the bound
states. Their contribution to the to the quadratic Hamiltonian (eq.(43)) is
\be
H^{(2)}\approx \sum_{k,i}\left(\frac{1}{2}\,P^{i}_{k}\,P^{i}_{- k}+
\frac{1}{2}\,\omega_{k}\,Q^{i}_{k}\,Q^{i}_{- k}\right).
\ee

We therefore have a collection of simple harmonic oscillators, labeled
by the integer $n$, with $k=2\,\pi\,n/p^{+}$, and $i=1,2$ ($D=2$).
Each mode contributes an amount 
\be
\sqrt{\omega_{k}}=u\,k=\frac{2\,\pi\,u}{p^{+}}\,n
\ee
where $u$ is a constant that can be read off from eq.(66)
 and $n$ is taken to be 
  small compared to $N_{0}$. We recall that the Hamiltonian
is the light cone variable $p^{-}$; and also the total transverse
momentum is zero because of the periodic boundary conditions on the
world sheet. The squared mass of the n'th excited state is then given by
\be
M_{n}^{2}=p^{+}\,p^{-}=2\,\pi\,u\,n+ M_{0}^{2}.
\ee
The zero point contribution  $M_{0}^{2}$, which we shall not try to calculate
here, is the sum of the renormalized
$E_{t}$  and the additional term gotten by normal ordering (72).
The above equation tells us that the spectrum is that of a string
with linear trajectories, where the slope $\alpha'$ is given by
\be
\alpha'=2\,\pi\,u=\frac{3^{1/2}\,\pi^{2}\,\bar{g}^{2}\,\rho_{0}^{2}\,
(1-\rho_{0})^{5/2}\,W^{2}(\rho_{0})}{4\,|\ln(\rho_{0})|^{3/2}}.
\ee
We therefore get a positive non-zero slope for a $\rho_{0}$ in the
allowed range $0\leq \rho_{0}\leq 1$, but different from zero or one.
The solution found in section 5  satisfies this condition.
Of course, the highly excited states of the string with mass squared
of the order of $N_{0}\,\alpha'$ merge with the continuum and the string
 picture breaks down. But as we stressed earlier, we can push this
transition region as high up as we wish by taking $N_{0}$ arbitrarily
large.

Next, we turn our attention to the case $D=4$. In this case, we need a more
extensive tuning of the parameters of the model in order to cleanly seperate
the string states from the continuum that starts at  $\bar{\omega}=1$. 
 Just as in the case $D=2$, $\bar{\omega}$ has to be small
compared to unity, and  therefore, we again
we look for solutions
 of the consistency equation (51) for small values of  $\bar{\omega}$
 by expanding $\eta\,
\rho_{0}/ F$ to second order in $k\,a$ as in eq.(60), and  the 
integral in (54) to first order in $\bar{\omega}$. The result is
\be
\bar{\omega}\approx 6\,\frac{1-\rho_{0}}{\rho_{0}^{2}\,|\ln(\rho_{0})|}
\,\ln\left(\frac{F\,\Lambda^{2}}{\lambda_{0}}\right)\,(k\,a)^{2}.
\ee
Since  $\bar{\omega}\ll 1$,  it follows that
$$
a^{2}\,\ln\left(\frac{\Lambda^{2}}{\lambda_{0}}\right)
$$
should be small. Or using (17), we have a more precise relation
\be
\ln\left(\frac{\Lambda^{2}}{\lambda_{0}}\right)/N_{0}^{2}=\nu \ll 1.
\ee

Let us try to make clear which parameters tend to infinity which of
them stay finite. Both $\lambda$ and $N_{0}$ tend to infinity, subject
to the above relation. 
 As $N_{0}\rightarrow \infty$, the two cutoff
 parameters $\Lambda$ and $N_{0}$ have to be corrolated:
\be
\Lambda^{2}\rightarrow \lambda_{0}\,\exp\left(\nu\,N_{0}\right)
\ee

In contrast, in this limit, $\nu$, although small compared to unity,
stays fixed and finite. Similarly, the remaining parameter $\lambda_{0}$
is also fixed and finite. It can be expressed in terms of the slope parameter
$\alpha'$ and $\nu$. Recalling that 
$\alpha'=2\,\pi\,u$, with $u$ given by eq.(73), we have,
\bq
\alpha'&=& 2\,\pi\,\left(6\,\frac{1-\rho_{0}}
{\rho_{0}^{2}\,|\ln(\rho_{0})|}\,\ln\left(\frac{F\,\Lambda^{2}}{\lambda_{0}}
\right)\right)^{1/2}\,\frac{p^{+}}{N_{0}}\,\lambda_{0}\nonumber\\
&\rightarrow& 2\,\pi\,\left(6\,\frac{1-\rho_{0}}
{\rho_{0}^{2}\,|\ln(\rho_{0})|}\,\nu \right)^{1/2}\,p^{+}\,\lambda_{0}.
\eq
This then the equation for the slope. Conversely, one can solve for
$\lambda_{0}$ in terms of $\alpha'$, $\rho_{0}$ and $\nu$. To sum it up,
with the tuning of the parameters described above, we have a string with
linear trajectories, whose low lying states are well seperated from the
continuum.

So far, expanding $H_{p}$ (eq.(41)) to second order around the classical
solution $\phi_{0}$, we have calculated only the free part
of the string Hamiltonian. This is all that is needed to determine
the spectrum.
 The interaction terms, which are cubic and
quartic, can be calculated from $H_{p}$. We shall not carry out this
calculation in this article.

\vskip 9pt

\noindent{\bf 9. Conclusions}

\vskip 9pt

In the present work, we have applied the mean field approximation to the
to the second quantized world sheet field theory developed in [1]. In this
approximation, we have found a non-trivial solution for the ground
state, and showed how to renormalize it. These results agree with those
of [1] based on the variational method, but the mean field method made it
also possible to compute fluctuations around this background. To
quadratic order, in addition to a continuum, a new spectrum of bound
states emerged. These bound states generate a bosonic string with linear
trajectories. We have shown that, by a suitable adjustment of the
parameters, it is possible to seperate the string states from the
continuum.

The $\phi^{3}$ theory on which the present work is based is an attractive
toy model for mainly its simplicity. It also shares the desirable feature of
asymptotic freedom with non-abelian gauge theories, but it is clearly
not a physical theory. The next target of research should be non-abelian
gauge theories; already some initial attempts were made in this direction
[14,15]. In analogy to what was done for the $\phi^{3}$ theory in [1],
we hope to develop the second quantized world sheet field theory for
gauge theories in various dimensions. It should then be relatively
straightforward to apply the mean field method to the resulting models and
see what comes out.

\vskip 9pt

\noindent{\bf Acknowledgement}

\vskip 9pt
This work was supported in part by the Director, Office of Science,
 Office of High Energy  Physics, of the U.S. Department of Energy under
 Contract DE-AC02-05CH11231.

\newpage

\noindent{\bf References}

\vskip 9pt

\begin{enumerate}

\item K.Bardakci, JHEP {\bf 0810} (2008) 056, arXiv: 0808.2959.
\item K.Bardakci and C.B.Thorn, Nucl.Phys. {\bf B 626} (2002) 287,
hep-th/0110301.
\item K.Bardakci and C.B.Thorn, Nucl.Phys. {\bf B 652} (2003) 196,
hep-th/0206205.
\item K.Bardakci, Nucl.Phys. {\bf B 715} (2005) 141, hep-th/0501107.
\item K.Bardakci, JHEP {\bf 0807} (2008) 057, arXiv:0804.1329.
\item H.P.Nielsen and P.Olesen, Phys.Lett. {\bf B 32} (1970) 203.
\item B.Sakita and M.A.Virasoro, Phys.Rev.Lett. {\bf 24} (1970) 1146.
\item G.'t Hooft, Nucl.Phys. {\bf B 72} (1974) 461.
\item A.Casher, Phys.Rev. {\bf D 14} (1976) 452.
\item R.Giles and C.B.Thorn, Phys.Rev. {\bf D 16} (1977) 366.
\item T.Banks, W.Fischler, S.H.Shenker and L.Susskind, Phys.Rev. {\bf
D 55} (1997) 5112, hep-th/9610043.
\item L.Susskind, hep-th/9704080.
\item K.Bardakci, Nucl.Phys. {\bf B 667} (2004) 354, hep-th/0308197.
\item C.B.Thorn, Nucl.Phys. {B 637} (2002) 272, hep-th/0203167, 
 S.Gudmundsson, C.B.Thorn and T.A.Tran, Nucl.Phys. {\bf B 649}
(2003), hep-th/0209102.
\item C.B.Thorn and T.A.Tran, Nucl.Phys. {\bf B 677} (2004) 289,
hep-th/0307203.

\end{enumerate}

\end{document}